\RequirePackage[]{graphicx} \graphicspath{{./Figures/}{./Graphs/}{./Drawings/}}

\documentclass[]{arxsub}

\usepackage[utf8]{inputenc}
\usepackage{graphicx}
\usepackage{environ}
\usepackage{listings}
\usepackage{booktabs}
\usepackage{blindtext}
\usepackage{url}
\usepackage{siunitx}

\usepackage[english]{babel}
\usepackage{csquotes}
\usepackage[
  backend=biber,
  natbib = true,
  style=ext-numeric-comp,
  sorting=none,
  minbibnames=3, maxbibnames=6,
  uniquename=false, uniquelist=false,
  giveninits=true, terseinits,
  articlein=false, innamebeforetitle=true,
  isbn=false,
  urldate=long, dateabbrev=false,
  autocite=superscript,
]{biblatex}

\DeclareNameAlias{sortname}{family-given}

\DeclareNameAlias{author}{sortname}
\DeclareNameAlias{editor}{sortname}
\DeclareNameAlias{translator}{sortname}

\DeclareDelimAlias{finalnamedelim}{multinamedelim}

\DeclareFieldFormat
  [article,inbook,incollection,inproceedings,patent,thesis,unpublished]
  {title}{#1}

\renewbibmacro*{journal+issuetitle}{\usebibmacro{journal}\setunit*{\addspace}\iffieldundef{series}
    {}
    {\newunit
     \printfield{series}\setunit{\addspace}}\usebibmacro{issue+date}\setunit{\addsemicolon\addspace}\usebibmacro{volume+number+eid}\setunit{\addcolon\space}\usebibmacro{issue}}

\DeclareFieldFormat[article,periodical]{number}{\mkbibparens{#1}}

\renewbibmacro*{issue+date}{\usebibmacro{date}}

\renewbibmacro*{pubinstorg+location+date}[1]{\printlist{location}\iflistundef{#1}
    {\setunit*{\addsemicolon\space}}
    {\setunit*{\addcolon\space}}\printlist{#1}\setunit*{\addsemicolon\space}\usebibmacro{date}\newunit}

\DeclareFieldFormat{pages}{#1}

\DeclareFieldFormat{doi}{doi\addcolon\space
  \ifhyperref
    {\href{https://doi.org/#1}{\nolinkurl{#1}}}
    {\nolinkurl{#1}}}

\AtEveryBibitem{\clearfield{number}\clearfield{month}\clearfield{day}\clearfield{urlyear}\clearfield{urlmonth}\clearfield{issn}\clearfield{series}\clearfield{pagetotal}\clearfield{eprintclass}\ifentrytype{article}{
	\iffieldundef{volume}{}{\clearfield{doi}}\clearfield{eprint}\clearfield{url}}{}\ifentrytype{book}{\clearfield{eprint}\clearfield{doi}\clearfield{url}}{\clearname{editor}}
\ifentrytype{inproceedings}{\clearfield{eprint}\clearfield{url}\clearfield{doi}\clearfield{publisher}\iffieldundef{booktitle}{}{\clearfield{eventtitle}}}{}\iffieldundef{eprint}{}{\clearfield{url}}
\iffieldundef{doi}{}{\clearfield{url}}
}

\usepackage{xpatch}

\makeatletter
\patchcmd\blx@bblinput{\blx@blxinit}
                      {\blx@blxinit
                      }{}{\fail}
\makeatother

\DeclareMathOperator*{\argmin}{arg\,min}

\newcommand{\figref}[2][]{\hyperref[#2]{Figure~\ref*{#2}#1}}
\newcommand{\tblref}[2][]{\hyperref[#2]{Table~\ref*{#2}#1}}
\renewcommand{\eqref}[1]{\hyperref[#1]{Eq.~\ref*{#1}}}

\newcommand{\bvec}[1]{\mathbf{#1}}

\newcommand{\norm}[1]{\left\lVert#1\right\rVert}

\providecommand{\doi}{}
\renewcommand{\doi}[1]{\href{https://doi.org/#1}{#1}}

\newcommand{\preSubmission}{Part of this work has been presented at the ISMRM Annual Conference 2022 (London).}

\providecommand{\del}[2][]{}
\providecommand{\dels}[2][]{}
\providecommand{\printfunding}{}

\title{Self-supervised learning for improved calibrationless radial MRI with NLINV-Net}

\corres{Moritz Blumenthal, Graz University of Technology,
Institute of Biomedical Imaging, Stremayrgasse~16/3, 8010~Graz, AUSTRIA, \email{blumenthal@tugraz.at}}

\author[1,2]{Moritz Blumenthal}{}
\author[1]{Chiara Fantinato}{}
\author[2,3,4]{Christina Unterberg-Buchwald}{}
\author[5]{Markus Haltmeier}{}
\author[6]{Xiaoqing Wang}
\author[1,2,4,7]{Martin Uecker}{}
\authormark{M. Blumenthal \textsc{et al.}}

\address[1]{Institute of Biomedical Imaging, Graz University of Technology, Graz, Austria}
\address[2]{Institute for Diagnostic and Interventional Radiology, University Medical Center Göttingen, Göttingen, Germany}
\address[3]{Clinic for Cardiology and Pneumology, University Medical Center Göttingen, Göttingen, Germany}
\address[4]{DZHK (German Centre for Cardiovascular Research), partner site Lower Saxony}
\address[5]{Department of Mathematics, University of Innsbruck, Innsbruck, Austria}
\address[6]{Department of Radiology, Harvard Medical School, Boston, Massachusetts, USA}
\address[7]{BioTechMed-Graz, Graz, Austria}


\articletype{Research Article}

\finfo{
This work was funded by the "Niedersächsisches Vorab" funding line of the Volkswagen Foundation
and it was funded in part by NIH under grant U24EB029240.
This work was supported by DZHK (German Centre for Cardiovascular Research) funding code: 81Z0300115.
The authors gratefully acknowledge the computing time granted by the Resource Allocation Board and provided on the supercomputer Lise and Emmy at NHR@ZIB and NHR@Göttingen as part of the NHR infrastructure.
}

\abstract{
\section{Purpose}
To develop a neural network architecture for improved calibrationless reconstruction of 
radial data when no ground truth is available for training. 
\section{Methods}
NLINV-Net is a model-based neural network architecture that directly estimates images and coil sensitivities
from (radial) k-space data via non-linear inversion (NLINV). Combined with a training strategy using self-supervision via data undersampling (SSDU), it can be used for imaging problems
where no  ground truth reconstructions are available. We validated the method for (1) real-time cardiac imaging 
and (2) single-shot subspace-based quantitative T1 mapping.
Furthermore, region-optimized virtual (ROVir) coils were used to suppress artifacts stemming from outside the FOV
and to focus the k-space based SSDU loss on the region of interest.
NLINV-Net based reconstructions were compared with conventional NLINV and PI-CS (parallel imaging + compressed sensing) reconstruction
and the effect of the region-optimized virtual coils and the type of training loss was evaluated
qualitatively.
\section{Results}
NLINV-Net based reconstructions contain significantly less noise than the NLINV-based counterpart.
ROVir coils effectively suppress streakings which are not suppressed by the neural networks while
the ROVir-based focussed loss leads to visually sharper time series for the movement of the myocardial wall
in cardiac real-time imaging.
For quantitative imaging, T1-maps reconstructed using NLINV-Net show similar quality as PI-CS reconstructions, 
but NLINV-Net does not require slice-specific tuning of the regularization parameter.
\section{Conclusion}
NLINV-Net is a versatile tool for calibrationless imaging which can be used in
challenging imaging scenarios where a ground truth is not available.
}

\begin{document}

\maketitle

\section{Introduction}

After the success of parallel imaging (PI) \cite{Sodickson_Magn.Reson.Med._1997,Pruessmann_Magn.Reson.Med._1999} and compressed sensing (CS) \cite{Block_Magn.Reson.Med._2007,Lustig_Magn.Reson.Med._2007}, deep learning based methods have further improved MRI reconstruction from undersampled data in recent years.
So-called physics-based or model-based network architectures \cite{Hammernik_Magn.Reson.Med._2017,Aggarwal_IEEETrans.Med.Imag._2019,Hammernik_Magn.Reson.Med._2021} build up on the classical PI-CS reconstruction by replacing the usually hand-crafted regularization terms by learnable network structures such as convolutional neural networks (CNNs).
If a large database with fully-sampled k-space data, such as the fastMRI database \cite{Zbontar_arXiv_2019}, is available
for training, physics-based neural networks can outperform classical PI-CS reconstruction \cite{Muckley_IEEETrans.Med.Imag._2021,Hammernik_IEEESignalProcess.Mag._2023}.
However, for many applications and specific MRI sequences large databases are not currently available, and it may be impractical or generally too expensive to acquire them for all imaging scenarios. In some cases, it may even be impossible to 
acquire fully-sampled ground truth data.
For example in cardiac MRI, supervised training usually requires breathholds and/or ECG gating to acquire ground truth data to train the reconstruction networks \cite{Schlemper_IEEETrans.Med.Imag._2018,Sandino_Magn.Reson.Med._2020,Kuestner_Sci.Rep._2020,Huang_Med.ImageAnal._2021,Kofler_Med.Phys._2021}.
For applications in cardiac real-time MRI this requires a restriction of the training dataset to simulations \cite{Hauptmann_Magn.Reson.Med._2019} or real-time scans where additionally ground truth references can be generated via breathholds and/or ECG gating \cite{Kleineisel_Magn.Reson.Med._2022}. 
Alternatively, if the goal of a neural network is the acceleration of online reconstruction rather than enhancement of image quality, time-consuming offline reconstruction may be used as reference \cite{Shen_NMRBiomed._2021}.

Recently, deep-learning methods have been developed, which improve reconstructions in settings where only undersampled k-space data is available.
Methods such as the deep image prior \cite{Yoo_IEEETrans.Med.Imag._2021,Hamilton_Magn.Reson.Med._2024}, zero-shot learning \cite{Yaman_Proc.Intl.Soc.Mag.Reson.Med._2022,Demirel_202345thAnnu.Int.Conf.IEEEEng.Med.Biol.Soc.EMBC_2023}, or RAKI \cite{Akcakaya_Magn.Reson.Med._2018} only make use of the specific undersampled dataset to be reconstructed by learning correlations within this dataset.
In contrast, self-supervision via data undersampling (SSDU) \cite{Yaman_Magn.Reson.Med._2020} can be used to learn from a large database of undersampled k-space data.
Model-based networks trained by SSDU on undersampled data have been shown to perform similarly well as the same network trained on fully-sampled ground truth data. 
A theoretical investigation \cite{Millard_IEEETrans.Comput.Imag._2023} interprets SSDU as a form of the Noisier2Noise \cite{Moran__2020} justifying that SSDU is equivalent to training on fully-sampled data with some restrictions on the sampling pattern.

While SSDU is an ideal candidate for enhancing image reconstruction for non-Cartesian cardiac MRI, prior publications \cite{Yaman__2021,MartinGonzalez__2021,Acar__2021} mostly focus on Cartesian data.
In this work, we explore the use of SSDU for radial acquisition schemes.
As radial trajectories sample a complementary mix of high and low frequency regions of k-space with each excitation, radial sampling is more robust to motion compared to Cartesian sampling and allows updating of the image content with each acquired spoke. For those reasons, radial sampling is ideally suited for dynamic imaging.
On the other side, reconstruction from radial data is sensitive to modeling inconsistencies caused by gradient delays or chemical shift \cite{Feng_J.Magn.Reson.Imaging_2022}. 
As the accurate mathematical modeling is a key requirement for the SSDU training strategy, these inconsistencies may hinder the application of self-consistency learning methods.
We stress that due to such physical effects which only appear for non-Cartesian sampling, simulation of non-Cartesian data from fully sampled Cartesian acquisitions often results in an inverse crime \cite{Colton__2013}.
Other problematic aspects which could arise when synthesizing
data are neglecting noise, trajectory errors, or discretization errors.
In this work, we investigate self-supervised learning for reconstruction of radial data, and apply it in two scenarios where fully-sampled ground-truth data is not available: Real-time cardiac MRI reconstructions and quantitative cardiac T1 mapping.

Most model-based network architectures need pre-estimated coil sensitivity maps, which are more difficult to obtain in the setting of strongly undersampled non-Cartesian datasets. For example, ESPIRiT \cite{Uecker_Magn.Reson.Med._2014} requires a fully-sampled Cartesian auto-calibration region, which needs to be recovered in a preprocessing step using gridding, which is not optimal.
Joint estimation of coil sensitivity maps and image content has been shown to improve conventional reconstructions \cite{Ying_Magn.Reson.Med._2007,Uecker_Magn.Reson.Med._2008} as well as network based reconstructions \cite{Sriram_MICCAI_2020,Luo__2021,Arvinte_MICCAI_2021}, and has been combined recently with self-supervised learning strategies \cite{Yuyang_Magn.Reson.Med._2024}.
The NLINV \cite{Uecker_Magn.Reson.Med._2008} method naturally supports non-Cartesian data by formulating the joint estimation of the coil sensitivities and image as a regularized non-linear inverse problem where image and coil sensitivities are separated using the assumption that the coil sensitivities are smooth functions.
NLINV has been successfully used for reconstruction of radial real-time data \cite{Uecker_Magn.Reson.Med._2010,Uecker_NMRBiomed._2010,Zhang_J.Cardiov.Magn.Reson._2010}.
Moreover, NLINV can integrate temporal regularization for the coil sensitivities, allowing for a continuous update of the coil sensitivity maps during real-time cine MRI which can be beneficial in the presence of respiratory motion \cite{Blaimer_Top.Magn.Reson.Imaging_2004}.
In this work, we propose NLINV-Net, a model based neural network architecture inspired by NLINV which jointly estimates coil-sensitivity maps and image content.
Since NLINV-Net only requires k-space data as input, it is an ideal network architecture for the SSDU training strategy, yielding a fully self-sufficient reconstruction method trained on undersampled data.
The basis structure of NLINV-Net can be adapted to various imaging scenarios.
Here, we consider two different reconstruction tasks, namely cardiac real-time MRI with a spatio-temporal version of NLINV-Net called RT-NLINV-Net and quantitative T1-mapping based on linear subspace reconstruction via Subspace-NLINV-Net.
Linear subspace reconstructions \cite{Petzschner_Magn.Reson.Med._2011,Tamir_Magn.Reson.Med._2017} reduce the dimensionality of the reconstruction problem by constraining the time series into a low-dimensional subspace.
Quantitative parameter mapping is then possible in post-processing step \cite{huang_Magn.Reson.Med.2012,Pflister_Magn.Reson.Med._2019,Roeloffs_IEEETrans.Med.Imag._2020}.
Recently, supervised learning \cite{Blumenthal__2022b,Iyer__2022} and zero-shot learning \cite{Jun_Magn.Reson.Med._2024} methods for enhancing linear subspace constraint reconstruction have been proposed.

Due to the success of deep learning based reconstruction methods, much research has been devoted
to finding optimal loss functions such as $\ell^1$-loss, SSIM, $\perp$-loss \cite{Terpstra_Med.ImageAna._2022} or perceptual losses \cite{Ghodrati_Quant.ImagingMed.Surg._2019} for training.
As the loss for SSDU training is formulated in k-space, most of these techniques are not directly applicable.
In this work, we investigate some alternative techniques.
For cardiac imaging, the region of interest, i.e. the heart, usually just takes up a small part
of the acquired FOV, hence, a masking of the loss function would be desirable to focus the loss to this ROI.
While masking of the loss in image space is not directly possible for SSDU training, we propose
to use region optimized virtual (ROVir) coil compression \cite{Daeun_Magn.Reson.Med._2021} to focus the k-space loss function to the respective ROI.
Moreover, we investigate in the use of using $\ell^1$ and $\ell^2$ based losses for 
radial SSDU. 
To summarize, the contributions of this work are as follows:
\begin{enumerate}
	\item We propose NLINV-Net as a self-contained neural network structure which does not require calibration of coil sensitivities
		and can be directly applied to acquired non-Cartesian k-space data.
	\item We propose and evaluate a method for focussing the SSDU training loss on a specific region of 
		interest based on ROVir coil compression. 
	\item We evaluate the proposed method in two examples: real-time cardiac MRI and quantitative T1 mapping.
\end{enumerate}

\section{Methods}
\label{sec:methods}

MRI reconstruction is usually formulated as an inverse problem, where the measurement process of the k-space data $\bvec{y}$ is modeled by the SENSE \cite{Pruessmann_Magn.Reson.Med._1999} model
\begin{align}
	\bvec{y}=F(\bvec{x}) + \eta = \mathcal{PF}(\rho \odot \bvec{c}) + \bvec{\eta}.
	\label{eq:signal}
\end{align}
Here, $\rho \in \mathbb{C}^{N\times N}$ is the transversal magnetization $\bvec{c}\in \mathbb{C}^{N\times N\times N_c}$ are the coil sensitivity maps, $\odot$ is the Hadamard product multiplying each of the $N_c$ coil sensitivity maps pixel-wise with the image $\rho$, $\mathcal{F}$ is the multichannel Fourier transform and $\mathcal{P}$ is the projection to the (potentially non-Cartesian) sampling pattern.
$\bvec{\eta}$ is the measurement noise, which is usually assumed to be complex white Gaussian noise, which is - after a pre whitening step - uncorrelated and equalized between channels.
In many works, the coil sensitivity maps $\bvec{c}$ are assumed to be known before the reconstruction such that only the image is unknown and the non-linear forward model $F$ reduces to a linear model.
In this case, the unknown $\bvec{x}$ is only the image $\bvec{x}=\rho$.
More generally, we consider in this work the joint reconstruction of the image and coil sensitivities where the reconstruction of the unknown $\bvec{x}=\left(\rho,\bvec{c}\right)$ is formulated as a regularized non-linear inverse problem
\begin{align}
	\bvec{x} = \argmin_{\bvec{x}} \norm{F(\bvec{x}) - \bvec{y}}_2^2 + \mathcal{R}(\bvec{x})\;. \label{eq:op}
\end{align}

\subsection{Self-Supervision via Data Undersampling (SSDU)}

Model-based neural networks for MRI reconstruction can generally be formulated as a function $\mathtt{Net}(\bvec{y}; \bvec{\theta}, F)$ mapping the k-space data $\bvec{y}$ to the reconstruction $\bvec{x}=\left(\rho,\bvec{c}\right)$.
The network itself is parametrized with the weights $\bvec{\theta}$ which are learned during the training.
As the structure of a model-based network is usually motivated by an optimization algorithm solving \eqref{eq:op}, the network is further parametrized by the forward model $F$.
For supervised training, pairs of undersampled k-space data $\bvec{y}_i$ and reference reconstructions $\hat{\bvec{x}}_i$ are created, and the weights are optimized with respect to some loss function $L$ according to
\begin{align}
	\bvec{\theta}^*=\argmin_{\theta} \sum_i L(\hat{\bvec{x}}_i , \mathtt{Net}(\bvec{y}_i;\theta, F))\;.
\end{align}
We stress that, from the acquired k-space data $\bvec{y}$, the reconstruction $\bvec{x}=(\rho , \bvec{c})$ is only defined up to an arbitrary non-zero complex-valued function $g$, i.e. if $\bvec{x}=(\rho , \bvec{c})$ is a solution of \eqref{eq:op}, then $\bvec{x}'=(\rho \cdot g , \bvec{c} / g)$ is a solution, too.
If the sensitivities are estimated in the pre-processing, their scale can be fixed by setting their sum-of-squares to one or by using a body coil to define a unique reference.
For a joint estimation of the image and coil sensitivity maps, the non-uniqueness of the coil sensitivity maps can be respected by selecting an invariant loss function $L$, for example by comparing the product of the coil sensitivity maps and the image $\rho \odot \bvec{c}$, or the corresponding k-space data.

In contrast to fully-supervised training, SSDU \cite{Yaman_Magn.Reson.Med._2020} is a training method which only requires the undersampled k-space data $\bvec{y}$ to learn a reconstruction.
For SSDU, the available k-space data is further subsampled by splitting it into two disjoint subsets $\bvec{y}_\Theta=M_\Theta \bvec{y}$ and $\bvec{y}_\Lambda=M_\Lambda \bvec{y}$ using respective binary masks $M_{\Theta/\Lambda}$.
We define the respective forward models $F_{\Theta/\Lambda}(\bvec{x})=M_{\Theta/\Lambda}\mathcal{PF}(\rho \odot \bvec{c})$ for the subsampled k-space data.
The network is then trained to perform a reconstruction of the data $\bvec{y}_\Theta$ which is consistent with the k-space data of the other subset $\bvec{y}_\Lambda$, i.e.
\begin{align}
	\bvec{\theta}^*=\argmin_{\theta} \sum_i L(\bvec{y}_{\Lambda,i} , F_\Lambda(\mathtt{Net}(\bvec{y}_{\Theta,i};\theta, F_\Theta)))\;.
	\label{eq:loss_ss}
\end{align}
Motivated by prior experiments with radial data \cite{Blumenthal__2022}, we split the data in this work spoke-wise into disjoint subsets such that $\bvec{y}_\Theta$ contains $75\%$ of the acquired data.
The masks are randomly generated for each training sample and are regenerated for each epoch.

\begin{figure*}
	\includegraphics[width=\linewidth]{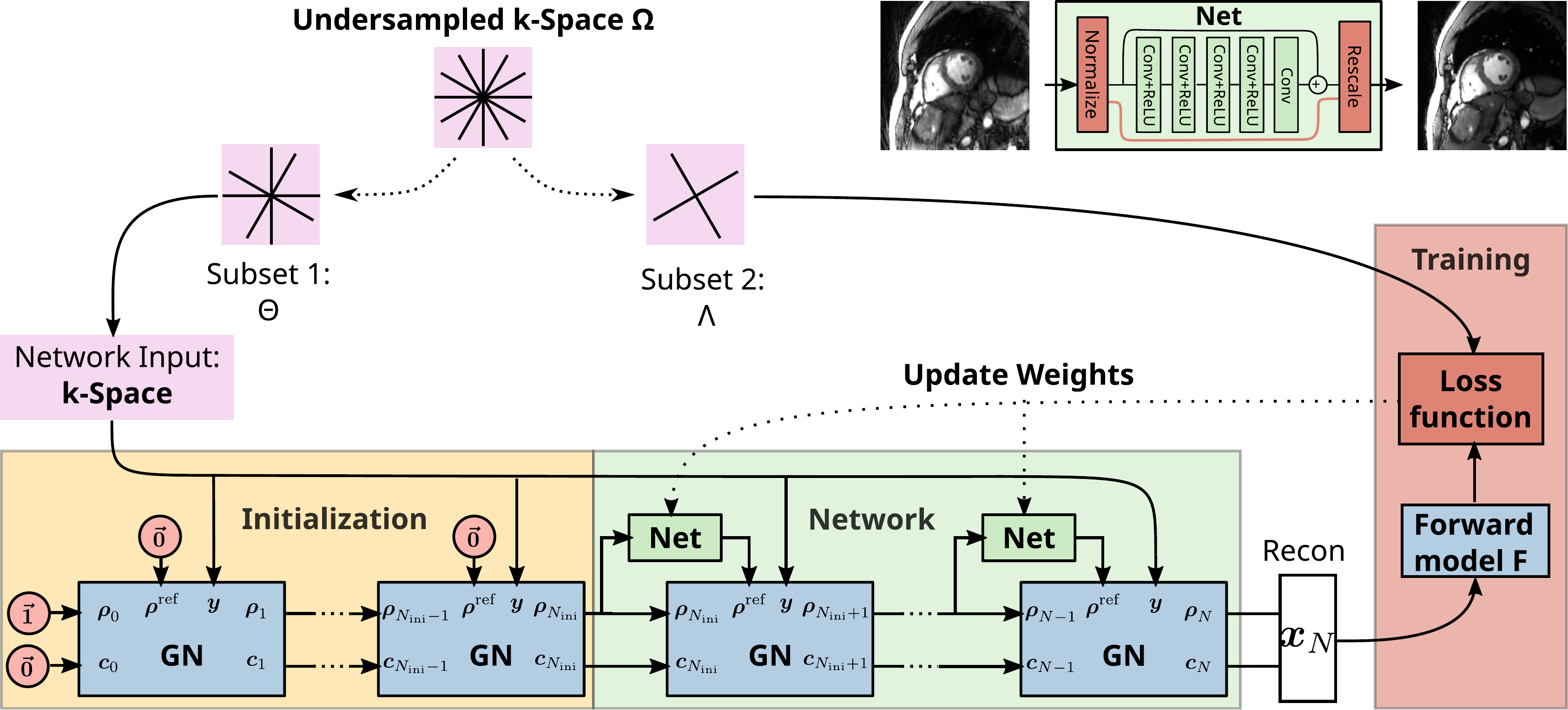}
	\caption{Schematic visualization of NLINV-Net with SSDU training strategy. NLINV-Net (bottom left) consists of an initialization phase (yellow) without network-based regularization and a second phase where the ResNet block (green) predicts a reference for the following Gauss-Newton data-consistency block (blue).
	The input of NLINV-Net is the raw k-space data (pink) and the output $\bvec{x}_N$ is the reconstruction and the estimated coil-sensitivity maps.
	For the SSDU training strategy, the radial k-space data is split spoke-wise into two subsets, from which one is feed into NLINV-Net and the other is used as reference data.}
	\label{fig:nlinvnet}
\end{figure*}

\subsection{NLINV-Net}

NLINV-Net is a model based neural network architecture which jointly estimates the image and the coil sensitivity maps based on NLINV.
In this section, we first describe NLINV-Net generally, before describing its variations for motion resolved and linear subspace based reconstructions in the subsequent sections.

NLINV solves the optimization problem in \eqref{eq:op} by the iteratively regularized Gauss-Newton method (IRGNM).
The regularization term $\mathcal{R}$ is split into an $\ell^2$-norm acting on the image $\rho$ and a weighted $\ell^2$-norm acting on the coil sensitivity maps $\bvec{c}$, i.e. $\mathcal{R}(\bvec{x}) = \alpha \norm{\rho}_2^2 + \alpha \norm{W^{-1}\bvec{c}}_2^2$ with the regularization parameter $\alpha$. To separate the image and the coil sensitivities, the weighting matrix $W^{-1}$ is chosen to correspond to the Sobolev norm to penalize high frequencies strongly \cite{Uecker_Magn.Reson.Med._2008}.
In each Gauss-Newton step, the non-linear forward model $F(\bvec{x})$ is linearized around the current estimate $\bvec{x}_n$, i.e. $F(\bvec{x})\approx F(\bvec{x}_n) + \left.\mathrm{D}F\right|_{\bvec{x}_n}(\bvec{x}-\bvec{x}_n)$ with the derivative $\left.\mathrm{D}F\right|_{\bvec{x}_n}$.
Inserting the linearized model in \eqref{eq:op} results in the linearized sub-problem
\begin{align}
	\bvec{x}_{n + 1} = \argmin_{\bvec{x} = (\rho, \bvec{c})}& \big\lVert \overbrace{F(\bvec{x}_n) + \left.\mathrm{D}F\right|_{\bvec{x}_n}\left(\bvec{x} - \bvec{x}_n\right)}^{\mathcal{PF}(\rho_n \odot \bvec{c} + \rho \odot \bvec{c}_n - \rho_n \odot \bvec{c}_n)} - \bvec{y}\big\rVert_2^2 \notag \\
	&+ \alpha_n \norm{\rho}_2 ^2 + \alpha_n \norm{W^{-1}\bvec{c}}_2^2 \,,
\end{align}
which yields a coupled update of the image $\rho$ and the coil sensitivity maps $\bvec{c}$. The optimization problem is solved by the conjugate gradient algorithm.
The regularization parameter $\alpha_n$ is halved after each Gauss-Newton step, while it is initialized with $\alpha_0=1$.

Following the idea of unrolling, NLINV-Net extends NLINV by a neural-network-based, learnable regularization, i.e. the $\ell^2$-regularization on the image is replaced by $\norm{\rho - \mathtt{CNN}(\rho_n; \theta)}_2^2$. Here, $\mathtt{CNN}$ is a denoising ResNet consisting of five 3x3 complex-valued convolutional layers with 64 channels and $\mathbb{C}$ReLU activation functions and $\theta$ denotes the network weights.
The input image $\rho_n$ is scaled to have a maximum magnitude of one and the output of $\mathtt{CNN}$ is rescaled to the original scale.
Since in the early iteration of the IRGNM the image $\rho_n$ is far away from the final reconstruction, we do not apply the network regularization in the first $N_{\mathrm{ini}}$ Gauss-Newton steps and use plain NLINV instead.
The number of initialization steps $N_{\mathrm{ini}}$ was selected based on visual inspection of the plain NLINV reconstructions.
After the initialization, the iterations for $N_\mathrm{ini} \le n < N$ of NLINV-Net can be expressed by:
\begin{align}
	\rho^{\mathrm{ref}}_{n+1} &=\mathtt{CNN}(\rho_n; \theta)
	\label{eq:NLINV-Net-Net}
	\\
	\bvec{x}_{n + 1} &= \argmin_{\bvec{x} = (\rho, \bvec{c})} \norm{ F(\bvec{x}_n) + \left.\mathrm{D}F\right|_{\bvec{x}_n}\left(\bvec{x} - \bvec{x}_n\right) - \bvec{y}}_2^2 \notag \\
	&\qquad\qquad + (\alpha_n + \lambda) \norm{\rho - \rho^{\mathrm{ref}}_{n+1}}_2^2 +(\alpha_n+\lambda_c) \norm{W^{-1}\bvec{c}}_2^2\;,
	\label{eq:NLINV-Net-GN}
\end{align}
where $\lambda>0$ and $\lambda_c>0$ are additional regularization parameters which are learned together with the weights of the $\mathtt{CNN}$ $\theta$.
After $N>N_{\mathrm{ini}}$ unrolled iterations, we obtain the final reconstruction $\bvec{x}_N(\bvec{y}; \theta, F)$ of NLINV-Net.
The network structure of NLINV-Net is visualized in \figref{fig:nlinvnet} where the optimization in \eqref{eq:NLINV-Net-GN} is visualized by blue boxes.
NLINV-Net is implemented using the deep-learning framework of BART \cite{Blumenthal_Magn.Reson.Med._2023} where similar to MoDL \cite{Aggarwal_IEEETrans.Med.Imag._2019} the conjugate gradient algorithm is used to compute the optimization in \eqref{eq:NLINV-Net-GN} and its derivatives.

\subsection{RT-NLINV-Net}

To exploit temporal correlation of consecutive frames for reconstruction of dynamic time-series, temporal regularization has been integrated in real-time (RT)-NLINV \cite{Uecker_NMRBiomed._2010} which penalizes the $\ell^2$ difference of the frame $\left(\rho^t, \bvec{c}^t\right)$ at time $t$ and the previous frame at time $t-1$.
The corresponding linearized sub problem reads
\begin{align}
	\bvec{x}_{n + 1}^t = \argmin_{\bvec{x} = (\rho, \bvec{c})} & \norm{ F(\bvec{x}_n^t) + \left.\mathrm{D}F\right|_{\bvec{x}_n^t}\left(\bvec{x} - \bvec{x}_n^t\right) - \bvec{y}^t}_2^2 \notag \\
	& \quad +
	\alpha_n \norm{\rho - d\rho_{N}^{t-1}}_2^2  + \alpha_n \norm{W^{-1}\left(\bvec{c}-d\bvec{c}^{t - 1}_{N}\right)}_2^2\,,
	\label{eq:RT-NLINV}
\end{align}
where $d=0.9$ is a temporal damping factor.

We integrate temporal regularization to RT-NLINV-Net by firstly using RT-NLINV with $N_{\mathrm{ini}}=6$ Gauss-Newton steps as initialization and secondly allowing the ResNet block to learn temporal regularization by extending its temporal receptive field.
For the latter, we slice $N_F=3$ consecutive frames $\left(\rho_n^{t - 1}, \rho_n^t, \rho_n^{t + 1} \right)^T$ from the reconstruction after every Gauss-Newton step and provide them channels to the ResNet to predict the center frame $\rho^{\mathrm{ref},t}_{n+1}$ as reference for the next Gauss-Newton step.
Hence, the network-regularized updates in the iterations $N_\mathrm{ini} \le n < N$ of NLINV-Net in \eqref{eq:NLINV-Net-Net}
and \eqref{eq:NLINV-Net-GN} is modified to
\begin{align}
	\rho^{\mathrm{ref},t}_{n+1} &=\mathtt{CNN}\left(\rho_n^{t - 1}, \rho_n^t, \rho_n^{t + 1}; \theta\right)
	\label{eq:RT-NLINV-Net-Net}
	\\
	\bvec{x}_{n + 1}^t &= \argmin_{\bvec{x}=(\rho, \bvec{c})}
	\norm{ F(\bvec{x}_n^t) + \left.\mathrm{D}F\right|_{\bvec{x}_n^t}\left(\bvec{x} - \bvec{x}_n^t\right) - \bvec{y}^t}_2^2 \notag \\
	&\hphantom{= \argmin_{\bvec{x}=(\rho, \bvec{c})}} + (\alpha_n + \lambda) \norm{\rho - \rho^{\mathrm{ref},t}_{n+1}}_2^2 \notag\\
	&\hphantom{= \argmin_{\bvec{x}=(\rho, \bvec{c})}} + (\alpha_n+\lambda_c) \norm{W^{-1}\left(\bvec{c}-d\bvec{c}^{t - 1}_{N_{\mathrm{ini}}}\right)}_2^2\;.
	\label{eq:RT-NLINV-Net-GN}
\end{align}
We use $N-N_{\mathrm{ini}}=3$ Gauss-Newton steps with network regularization.
While the temporal regularization in RT-NLINV is causal, the network based regularization also depends on data from future frames. 
Nevertheless, the sliding window approach allows principally for an online real-time reconstruction where the reconstruction is delayed by $(N-N_{\mathrm{ini}})(N_F-1)/2=3$ frames with respect to the currently acquired k-space data.
Every Gauss-Newton step with network regularization increases the delay by one frame similar to the growth of the receptive field in a CNN with every layer (c.f. Supplementary Information \figref{supfig:receptive_field}).

To exclude boundary effects and suppress the initialization phase of RT-NLINV, the first eight and last three frames are excluded from the self-supervised training loss.
To obtain reconstructed images corresponding to temporally smooth coil sensitivity maps, we apply a running average filter of size 5 on the estimated coil sensitivity maps before applying $F_\Lambda$ to compute the self-supervised loss in \eqref{eq:loss_ss}.

\subsection{Subspace-NLINV-Net}
Subspace-NLINV-Net is an adaption of NLINV-Net for linear subspace reconstruction, which we describe here for quantitative T1-mapping.
We consider data from an inversion recovery sequence with  a continuous radial FLASH readout.
The signal at the $j$-th excitation of this sequence follows the Look-Locker signal model \cite{Look_Rev.Sci.Instrum._1970,Deichmann_J.Magn.Reson._1992}
\begin{align}
	\rho^j=\rho_{\mathrm{SS}} - \left(\rho_{\mathrm{SS}} + \rho_{0}\right)\exp\left(-\frac{j \Delta t}{T_1^*}\right), \label{eq:signal_ll}
\end{align}
with the steady state magnetization $\rho_{\mathrm{SS}}$, the equilibrium magnetization $\rho_{0}$, the effective $T_1$-time $T_1^*$ and the time difference of two excitations $\Delta t=\mathrm{TR}$.
If all three parameters ($\rho_{\mathrm{SS}}$, $\rho_{0}$, $T_1^*$) are estimated, $T_1$ can be computed by
\begin{align}
	T_1=T_1^* \frac{\rho_{0}}{\rho_{\mathrm{SS}}}+2t_{\mathrm{d}}, \label{eq:corr}
\end{align}
where $t_{\mathrm{d}}$ is the delay from the inversion pulse to the first excitation \cite{Deichmann_Magn.Reson.Med._2005}.

We consider a linear subspace constrained reconstruction where the magnetization $\rho^j\approx\sum_{s=1}^{N_S}\mathcal{B}_{js}\bvec{a}_s$ is expressed by a linear combination of a few ($N_S=4$) subspace coefficients $\bvec{a}_s$.
The subspace basis $\mathcal{B}$ is computed by first generating a dictionary of signal evolutions from \eqref{eq:signal_ll} for 1000 different $T_1$-values varying linearly from 0.1 to 4s and 100 values for the effective flip angle $\alpha$ varying linearly from $2^\circ$ to $4^\circ$. The corresponding value for $T_1^*$ is computed via $1/T_1^* = 1/T_1 + 1/\mathrm{TR}\log[\cos(\alpha)]$ and $\rho_{\mathrm{SS}} = \rho_0 T_1^*/T_1$. Afterward, the subspace basis is computed by a PCA considering only the first $N_S=4$ principal components.
To cope with the cardiac motion, a subject specific temporal mask $\mathcal{D}$ selecting diastolic data is introduced, which discards data from other motion states by weighting them with zero.
The subject specific temporal mask does not influence the subspace basis so that the coefficient maps have the same physical interpretation across subjects and their correlations can be learned.
Both, the subspace basis $\mathcal{B}$ and the diastole mask $\mathcal{D}$ are integrated in the non-linear forward model such that we directly reconstruct the subspace coefficients $\bvec{a}$ and coil sensitivity maps $\bvec{c}$ in the diastolic phase, i.e. the forward model $F$ from \eqref{eq:op} reads
\begin{align}
	\bvec{y}=F\left(\bvec{x}\right) + \eta = \mathcal{PF}(\mathcal{D}\mathcal{B}\bvec{a} \odot \bvec{c}) + \bvec{\eta}.
\end{align}
Here, $\mathcal{DB}$ commutes with the Fourier transform $\mathcal{F}$ such that it can be merged with the sampling pattern $\mathcal{P}$.
This enables an efficient implementation of the matrix-vector product with $\mathcal{F}\mathcal{B}^H\mathcal{DP}\mathcal{B}\mathcal{F}$ required during the computation of the linearized sub-problem (\eqref{eq:NLINV-Net-GN}) by combining the subspace basis with Toeplitz embedding of the nuFFT \cite{Fessler_IEEETrans.SignalProcessing_2005,Uecker_Magn.Reson.Med._2010,Mani_Magn.Reson.Med._2015,Tamir_Magn.Reson.Med._2017}.
Due to this implementation, the computational cost is independent of the number of inversion times and no binning of spokes to temporal frames is required.
The extension of NLINV-Net to Subspace-NLINV-Net is straightforward by replacing the forward model.
We use $N=15$ Gauss-Newton steps for Subspace-NLINV-Net from which $N-N_{\mathrm{ini}}=3$ are regularized by the $\mathtt{CNN}$ which takes the $N_S=4$ coefficient maps as different channels.
After reconstruction of the coefficient maps, the signal evolution of \eqref{eq:signal_ll} is projected to the linear subspace and fitted pixel-wise to the coefficient maps to obtain parameter maps for $\rho_{\mathrm{SS}}$, $\rho_{0}$ and $T_1^*$. Finally, the $T_1$ map is obtained by \eqref{eq:corr}.

\begin{figure*}
	\includegraphics[width=\linewidth]{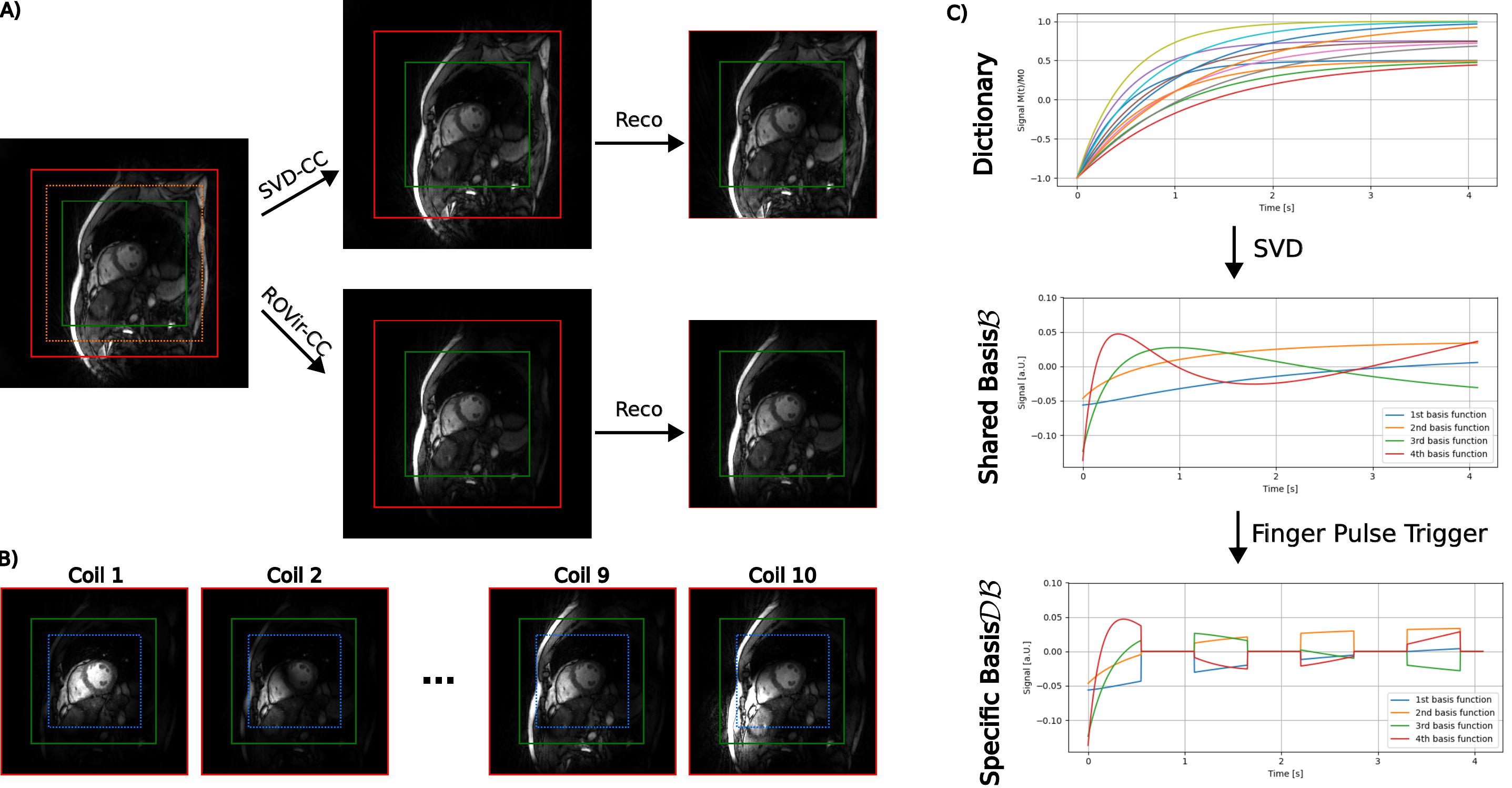}
	\caption{(A) Visualization of nominal FOV (green square) and reconstruction grid size (red square).
	ROVir-CC suppresses signal from outside the orange dotted square and maximizes signal from the nominal FOV leading to less modeling error from signal from outside the reconstruction grid.
	(B) Application of another ROVir-CC matrix focuses the signal from the blue dotted box in the first virtual coil image.
	(C) The shared subspace basis $\mathcal{B}$ is generated from a dictionary via SVD.
	A subject specific temporal pattern selecting only diastolic data is generated based on a finger pulse trigger.}
	\label{fig:rovir}
\end{figure*}

\subsection{Training Datasets and Preprocessing}

All MRI data has been acquired in previous studies with written informed consent and with approval of the local ethics committee on a Magnetom Skyra 3T (Siemens Healthineers, Erlangen).

For RT-NLINV-Net, k-space data from an interleaved 5-turn radial bSSFP (TE/TR=1.29/$\SI{2.58}{ms}$, FA=23°, FOV=$256 \times \SI{256}{mm^2}$, slice thickness: $\SI{6}{mm}$, matrix size: $160 \times 160$) sequence was used.
Each frame consists of 13 uniformly spaced spokes corresponding to a nominal undersampling factor of 19 and acquisition time of approximately $\SI{33}{ms}$ per frame.
After each frame the trajectory is rotated by $360^\circ/65$ such that the sampling repeats after 5 frames (turns).
This trajectory was introduced for real-time MRI in Reference \cite{Zhang_J.Cardiov.Magn.Reson._2010}.
The training dataset contains 22-27 slices with 127 frames ($\approx \SI{4}{s}$) of short-axis views from 40 volunteers.
After the estimation of gradient delays and coil compression described below, the data is split into shorter sequences consisting of 26 frames each, resulting in 3820 training samples.

The T1 inversion recovery training dataset contains three (apex, mid and base) slices of short-axis views from 43 volunteers.
Some volunteers have been scanned multiple times, resulting in total in 227 slices. The dataset is detailed in \cite{Kohler_SCMR_2020}.
After a single, non-selective inversion pulse, 1530 radial spokes with the 9th tiny golden angle were acquired with a continuous FLASH readout (TE/TR: 1.67/$\SI{2.67}{ms}$, FA=4°, FOV=$256 \times \SI{256}{mm^2}$, slice thickness: $\SI{8}{mm}$, matrix size: $256 \times 256$), resulting in a scan time of approximately $\SI{4}{s}$ per slice.
The initial inversion pulse was triggered in early diastole by a finger pulse signal after a delay of $\SI{0.1}{s}$.
The diastolic mask $\mathcal{D}$ was then generated based on the finger pulse signal to retrospectively discard data acquired in an interval from $\SI{0.4}{s}$ before to $\SI{0.1}{s}$ after the following trigger signals, resulting on average in about 800 spokes per slice.
For more details on the trigger, we refer to \cite{Wang_Brit.J.Radiol._2016}.

The last 65 (real-time) or 85 (T1) spokes of each dataset -- corresponding to the steady state in the inversion-recovery sequence -- were used for the estimation of gradient delays using RING~\cite{Rosenzweig_Magn.Reson.Med._2018a} and computation of the coil compression matrix.
We evaluate the effect of two different coil compression (CC) methods, namely SVD-based CC \cite{Buehrer_Magn.Reson.Med._2007,Huang_Magn.Reson.Imaging_2008} and ROVir-CC \cite{Daeun_Magn.Reson.Med._2021}.

Coil compression is often used to reduce the computational cost of MRI reconstructions by compressing k-space data $\bvec{y}_j$ of the $N_c$ acquired coils to the virtual k-space data $\tilde{\bvec{y}}_i = \sum_{j=1}^{\tilde{N}_c}\bvec{U}_{ij}\bvec{y}_j$ consisting of $\tilde{N}_c \le N_c$ virtual coils.
$\bvec{U}\in\mathbb{C}^{N_c\times N_c}$ is a unitary matrix such that the optimization objective in \eqref{eq:op} is invariant under this transform in the case of no compression ($\tilde{N}_c=N_c$).
Usually, the compression matrix $\bvec{U}$ is determined by an SVD such that it rotates most signal in the first virtual channels and only noise contained in the later channels is discarded by compression to $\tilde{N}_c < N_c$ virtual coils.
We refer to this method by SVD-CC and compress the data to $\tilde{N}_c=10$ virtual coils.

Since the radial data were acquired with twofold oversampling in readout direction, there is some freedom in choosing the FOV for the reconstruction grid, ranging from the nominal FOV to a grid increased by a factor of two in both directions.
As the chest in cardiac imaging usually exceeds the nominal FOV, a trade-off between computational costs for larger grids and boundary effects for too small grids often leads to 1.5 times enlarged grid for reconstruction.
In \figref{fig:rovir}A, we mark the boundaries of the reconstruction grid with a red square and the nominal FOV with a green square.

In contrast to SVD-CC, ROVir-CC \cite{Daeun_Magn.Reson.Med._2021} allows defining two regions and computes a coil compression matrix $\bvec{U}$ such that the signal stemming from the first region is maximized and the signal stemming from the second region is minimized in the first virtual coils.
For NLINV-Net, ROVir-CC can have two effects, i.e. first the signal from outside the reconstruction grid can be minimized which results in smaller modeling error and second, since the signal of a region of interest (ROI) is maximized also the self-supervised training loss (\eqref{eq:loss_ss}) is focussed on this ROI.
Hence, we use ROVir-CC to maximize signal from the nominal FOV while suppressing the signal from outside the 1.25-fold enlarged nominal FOV (c.f. \figref{fig:rovir}B).
To compute the ROVir-CC matrix, the k-space data is projected to the signal stemming from the respective regions by first computing the inverse nuFFT on a two-fold over-sampled low-resolution ($80\times80$) grid, second multiplication with the respective region-masks, and third a forward nuFFT to transform the signal back to the non-Cartesian k-space.
Based on these signals, we compute the coil compression matrix $\bvec{U}$ as suggested in \cite{Daeun_Magn.Reson.Med._2021} and employ the Gram-Schmidt algorithm to ensure that $\bvec{U}$ is indeed unitary. The data is again compressed to $\tilde{N}_c=10$ virtual coils.

Finally, we again use ROVir to rotate the $\tilde{N}_c=10$ virtual coils of both coil compression methods such that the signal stemming from the center 75\% of the nominal FOV is maximized, and the remaining signal is minimized in the first coils. 
As the coils are not further compressed, any optimization is invariant under this transform. However, the transform allows us to focus the self-supervised train loss to the center containing the heart by restricting the loss in \eqref{eq:loss_ss} to the k-space data of the first virtual coil.
We refer to this method as focussed loss instead of a global loss when we use k-space data from all virtual coils.

\subsection{Experiments}

For both datasets, NLINV-Net is trained from scratch using the SSDU loss function for eight combinations of preprocessing steps and loss functions. Namely, we vary the coil compression method (SVD or ROVir) and the training loss (global MSE, focussed MSE, global MAD and focussed MAD).
The networks are trained for $\SI{10}{h}$ (real-time) or 250 (T1) epochs with a mini-batch size of 16 taking about $\SI{10}{h}$ (real-time) or $\SI{7}{h}$ (T1) on a server equipped with 4 Nvidia A100 ($\SI{80}{GB}$ HBM2, SXM) GPUs.
Reconstruction with RT-NLINV-Net of the full time series (127~frames / $\SI{4}{s}$) takes about $\SI{50}{s}$ on a single Nvidia H100 GPU ($\SI{80}{GB}$ HBM3, SXM), reconstructing a slice of the T1 dataset takes about $\SI{4}{s}$ with Subspace-NLINV-Net.
For comparison to RT-NLINV-Net, we perform RT-NLINV reconstructions with SVD-CC and ROVir-CC.
We also provide comparisons to locally low-rank \cite{Trzasko__2011} PI-CS reconstruction, which is not real-time capable, and RT-NLINV-Net with varying temporal receptive field in Supplementary Information \figref{supfig:llr_frames}.
As reference for the subspace reconstruction, we use Subspace-NLINV with 12 Gauss-Newton steps.
Moreover, exploiting spatial correlations in the coefficient maps, we compute joint $\ell^1$-Wavelet regularized reconstructions \cite{Wang_Philos.Trans.R.Soc.A._2021}  for different regularization parameters using the fixed coils estimated with Subspace-NLINV.
 
\section{Results}
\label{sed:results}

\subsection{Real-Time Dataset}

\begin{figure*}
	\includegraphics[width=\linewidth]{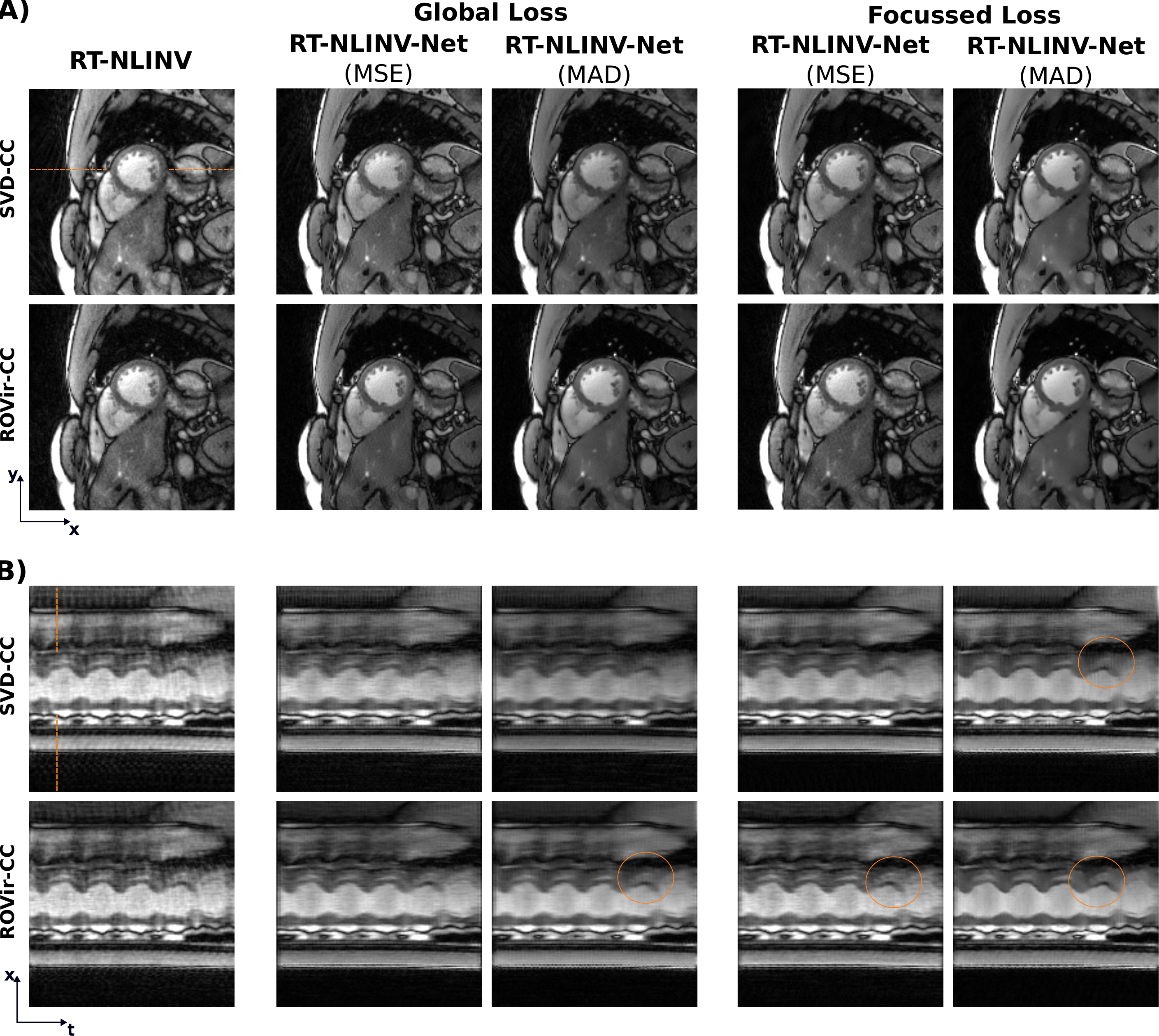}
	\caption{
		Example reconstructions shown in the X-Y (A) and Y-time (B) plane reconstructed by RT-NLINV and RT-NLINV-Net trained with different loss function. Dotted orange lines mark cross-sections shown in the other plane. Overall, RT-NLINV-Net denoises reconstructions compared to RT-NLINV while ROVir-CC is superior to SVD-CC, MAD-loss is superior to MSE-loss and focussed loss is superior to global loss.
		Orange circles mark temporal flickering in the cardiac region, which is most suppressed by RT-NLINV-Net trained with focussed MAD loss. 
	}
	\label{fig:rt1}
\end{figure*}

\begin{figure*}
	\includegraphics[width=\linewidth]{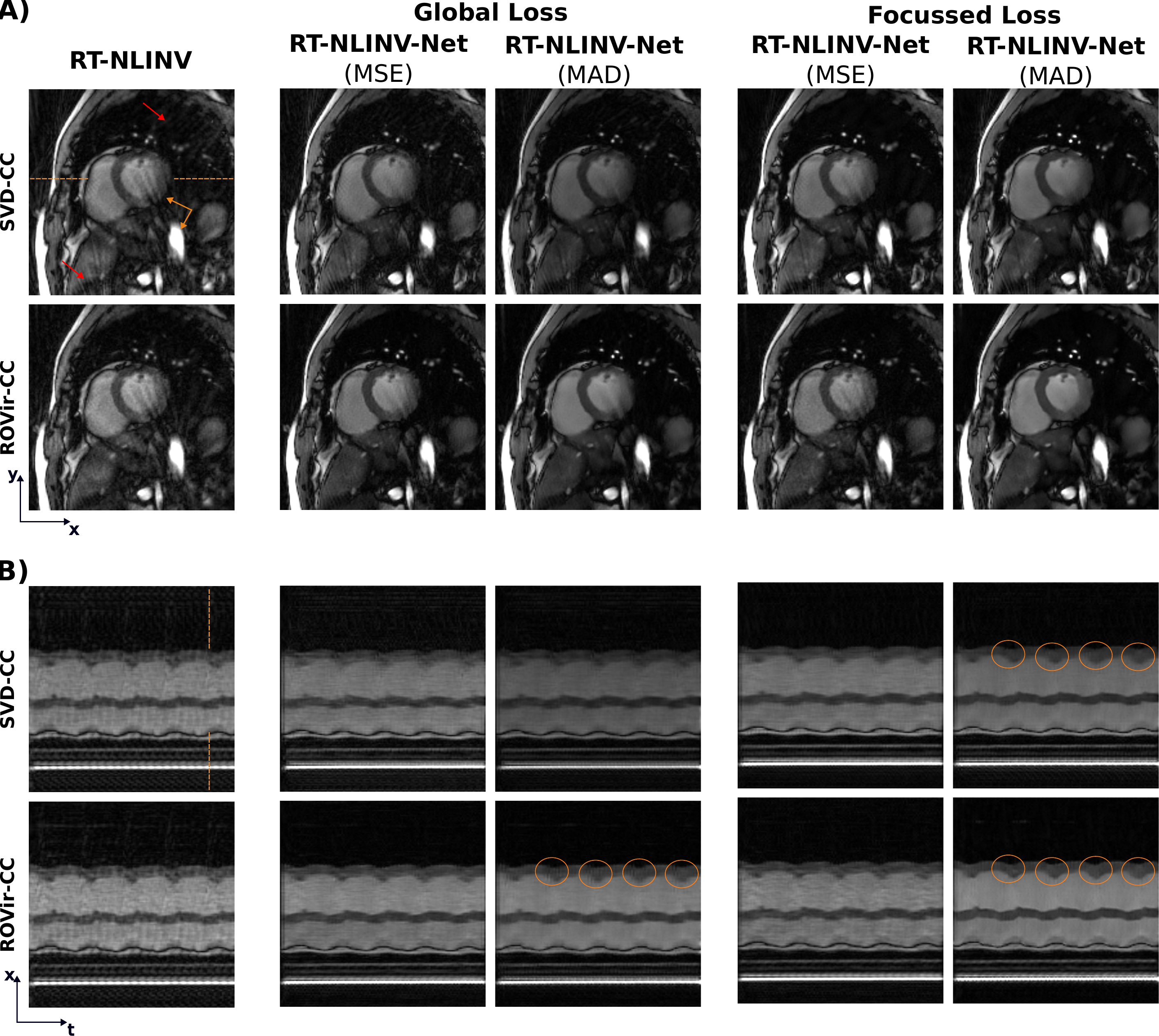}
	\caption{
		Example reconstructions for a different volunteer than in \figref{fig:rt1} with more streaking artifacts in the RT-NLINV reconstruction.
		Streakings originate from outside the FOV (red arrows) and pulsating blood (orange arrows).
		ROVir-CC reduces the streakings from outside the FOV while the streakings from the blood vessel are only suppressed by RT-NLINV-Net trained with a focussed MAD-loss.
		Orange circles mark flickering in the cardiac wall due to these streakings.
	}
	\label{fig:rt2}
\end{figure*}

To evaluate the effectiveness of the RT-NLINV-Net trained with the different loss functions qualitatively, we show in Figures \ref{fig:rt1} and \ref{fig:rt2} (and correspondingly Video S1 and Video S2)
example reconstructions using RT-NLINV and RT-NLINV-Net.
The datasets have been selected as representative examples for an RT-NLINV reconstruction mostly corrupted by noise (\figref{fig:rt1}) 
and for one affected by strong streaking artifacts (\figref{fig:rt2}). 

\begin{figure}
	\centering
	\includegraphics[width=.7\linewidth]{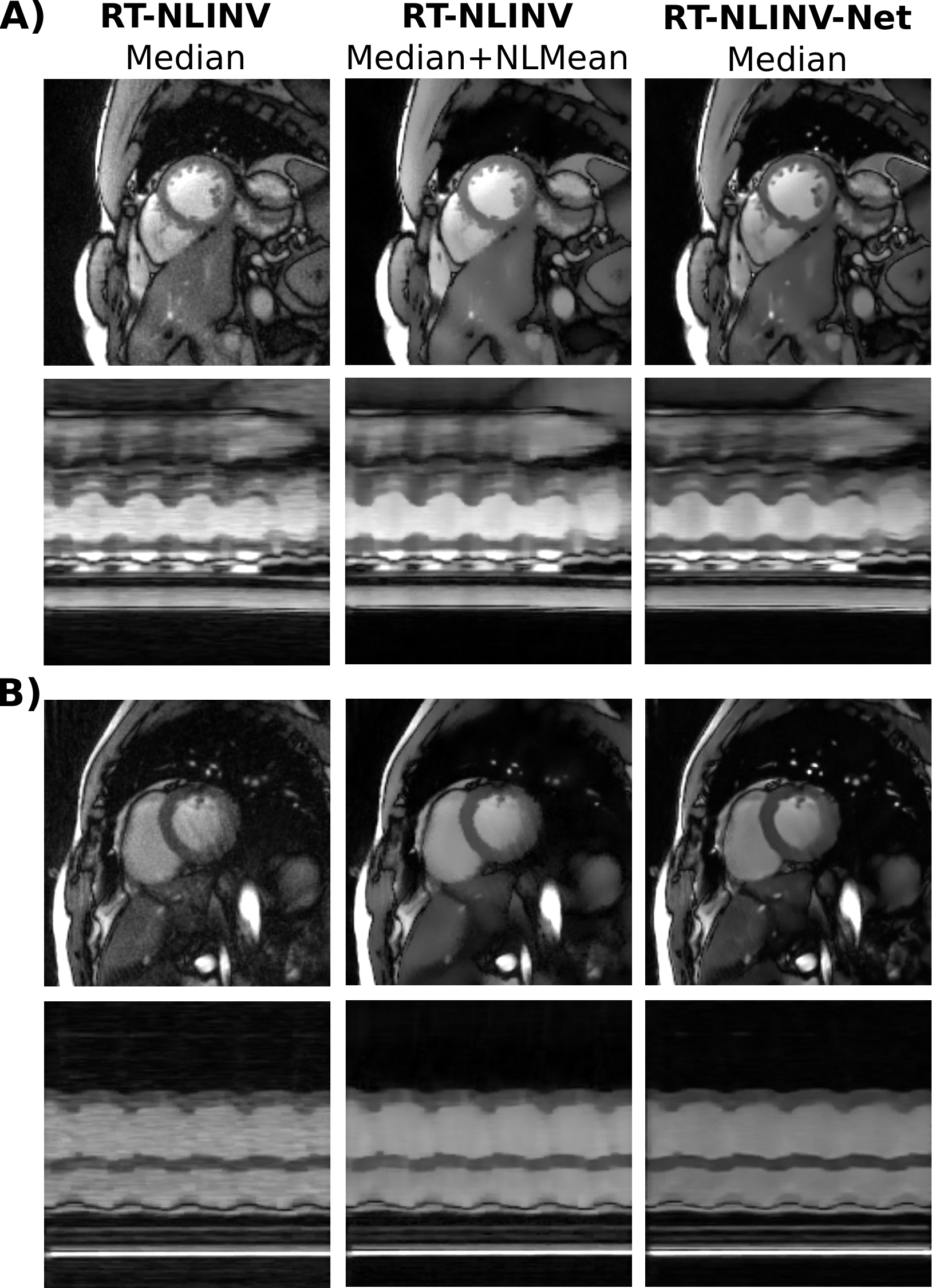}

	\raggedright
	\caption{
	Post-processed reconstructions of the example reconstructions in \figref{fig:rt1} (A) and \figref{fig:rt2} (B).
	Temporal median filtering removes the flickering visible in the plain reconstructions. 
	The non-local means filter removes residual noise in the RT-NLINV reconstructions but still provides some temporal blockiness compared to the median filtered RT-NLINV-Net reconstruction.
	In B, median and non-local means filtering reduces streaking artifacts significantly, while they are still stronger suppressed in the RT-NLINV-Net reconstruction.	
	}
	\label{fig:rt_post}
\end{figure}

For the reconstructed frame in \figref{fig:rt1}A, the choice of the coil compression method has a minor effect on the reconstruction quality, however, streaking artifacts outside the body originating 
from the subcutaneous fat are slightly stronger suppressed for ROVir-CC than for SVD-CC.
RT-NLINV-Net reconstructions trained with any loss function show less noise compared to the respective plain RT-NLINV reconstruction.
Comparing the RT-NLINV-Net reconstructions among each other, RT-NLINV-Net trained with the MAD-loss denoises the reconstruction stronger than the MSE counterpart and the focussed loss performs superior compared to the global counterpart.
Consequently, the best denoising is achieved with RT-NLINV-Net trained by a focussed MAD loss and applied on ROVir-CC data.
In \figref{fig:rt1}B, we present the temporal evolution of a horizontal slice through the left ventricle of the respective reconstructions in \figref{fig:rt1}A.
All reconstructions reveal some temporal flickering which is visible by periodical vertical stripes, which is especially visible in the RT-NLINV reconstructions.
This pattern originates from the interleaved nature of the radial sequence, which acquires every five frames the same radial spokes.
Usually, the pattern can be suppressed in post-processing via temporal median-filtering \cite{Uecker_NMRBiomed._2010} as presented in \figref{fig:rt_post}.
Comparing the different reconstructions, ROVir-CC reduces temporal flickering compared to SVD-CC, especially for the RT-NLINV-Net reconstructions. 
Generally, RT-NLINV-Net trained with a focussed loss reduces the flickering in the region of the heart where the combination of ROVir-CC and a focussed MAD loss almost completely removes flickering in the heart, including the region marked by an orange circle.

\figref{fig:rt2} shows reconstructions as in \figref{fig:rt1} but for an example dataset where the RT-NLINV reconstruction exhibits significant streaking artifacts which originate from outside the nominal FOV (red arrows) and a vessel inside the FOV (orange arrows).
The artifacts from outside the FOV are suppressed by ROVir-CC.
RT-NLINV-Net also suppresses those in the lung for SVD-CC.
The more relevant streakings in the heart are only suppressed by the reconstruction of RT-NLINV-Net trained by the focussed MAD loss and applied on ROVir-CC data.
Hence, this reconstruction performs overall best on this example. 
Beyond the mentioned streakings, the RT-NLINV-Net trained with MAD again provides a more denoised reconstructions than MSE loss and the focussed loss denoises slightly more than global loss.
The temporal evolution of a horizontal slice through the respective reconstructions is presented in \figref{fig:rt2}B.
Here, the streaking artifacts originating from the blood in the vessel manifest as flickering in the myocardial wall during the systole.
This flickering is strongly but not completely suppressed in the RT-NLINV-Net reconstruction trained with focussed MAD.

In \figref{fig:rt_post} and respectively in Video S3 and Video S4, we present post-processed versions of the reconstructions in \figref{fig:rt1} and \figref{fig:rt2}.
Temporal median filtering with a filter length of five frames corresponding to the five turns of the interleaved sequence has been applied to reduce the temporal flickering.
Further, non-local means filtering \cite{Buades_2005CVPR_2005} can be used to reduce the noise of the RT-NLINV reconstruction, but it is not helpful to the already denoised RT-NLINV-Net reconstruction.
While the post-processing improves the RT-NLINV reconstructions, the time series still exhibits some blockiness with staircasing effect in the separation of the left ventricle and the myocardial wall.
In contrast, the median-filtered reconstruction of RT-NLINV-Net shows a more continuous movement of the myocardial wall with clear edges separating it from the blood pool in the left ventricle.
While the post-processing reduces the streaking artifacts in the RT-NLINV reconstructions, they are still more suppressed in the median-filtered RT-NLINV-Net reconstruction.
For completeness, we include the coil sensitivity maps of the reconstructions in the supporting information.

\subsection{T1 Dataset}

\begin{figure*}
	\includegraphics[width=\linewidth]{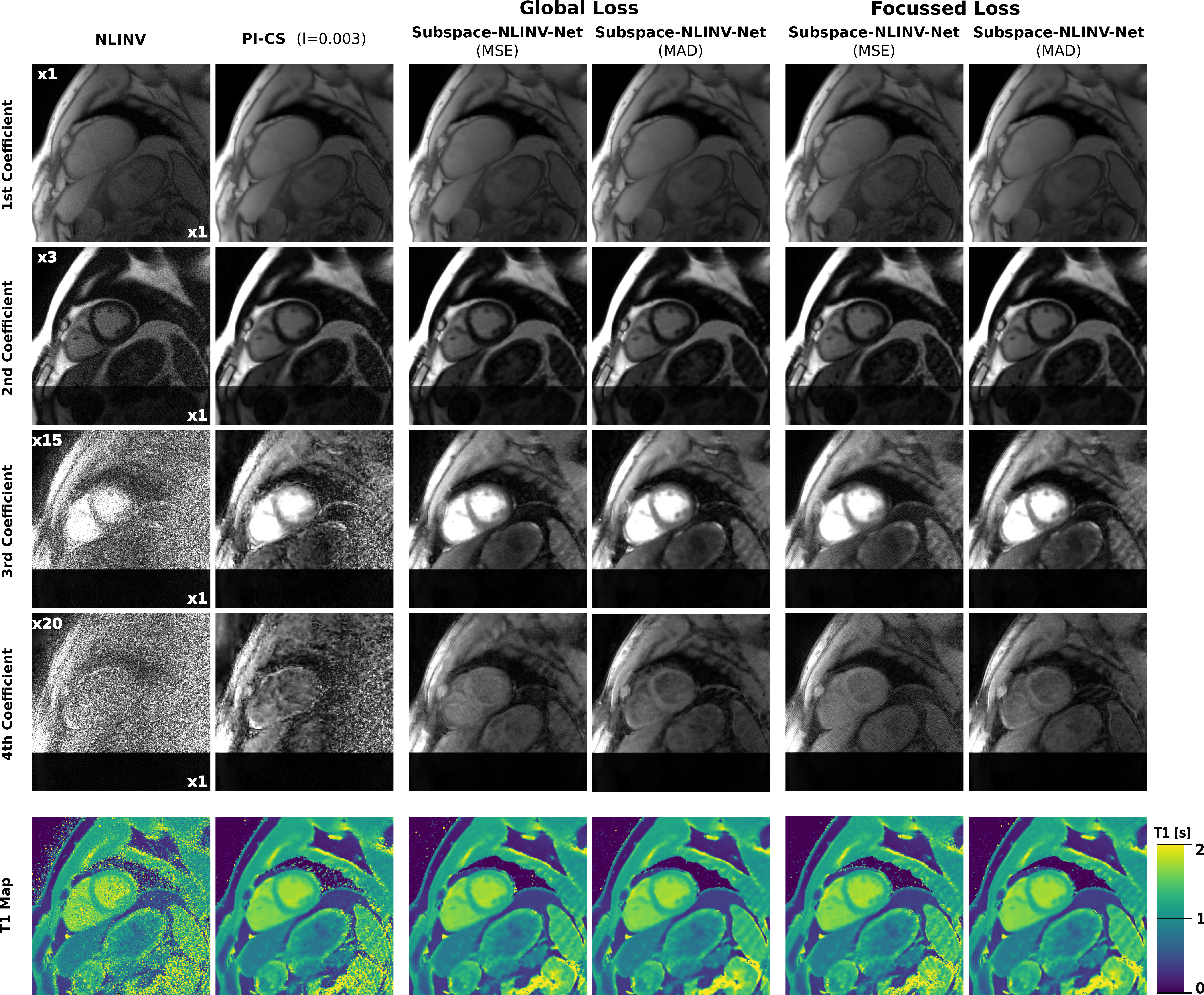}	
	\caption{Subspace reconstruction and corresponding $T_1$-maps of a (mid) short-axis slice with SVD-CC using NLINV, an $\ell^1$-regularized PI-CS reconstruction and Subspace-NLINV-Net reconstructions trained by different losses.
	The plain NLINV reconstruction is too noisy, resulting in a corrupted $T_1$-map.
	$\ell^1$-Wavelet regularization denoises the reconstruction while introducing wavelet artifacts in the third and fourth coefficient maps.
	NLINV-Net denoises the reconstructions for all training losses, but NLINV-Net trained with a focussed MSE introduces some high-frequency pattern.
	NLINV-Net with focussed MAD provides a slightly sharper reconstruction (c.f. papillary muscels).
	}
	\label{fig:coeff_svd}
\end{figure*}

\begin{figure*}
	\includegraphics[width=\linewidth]{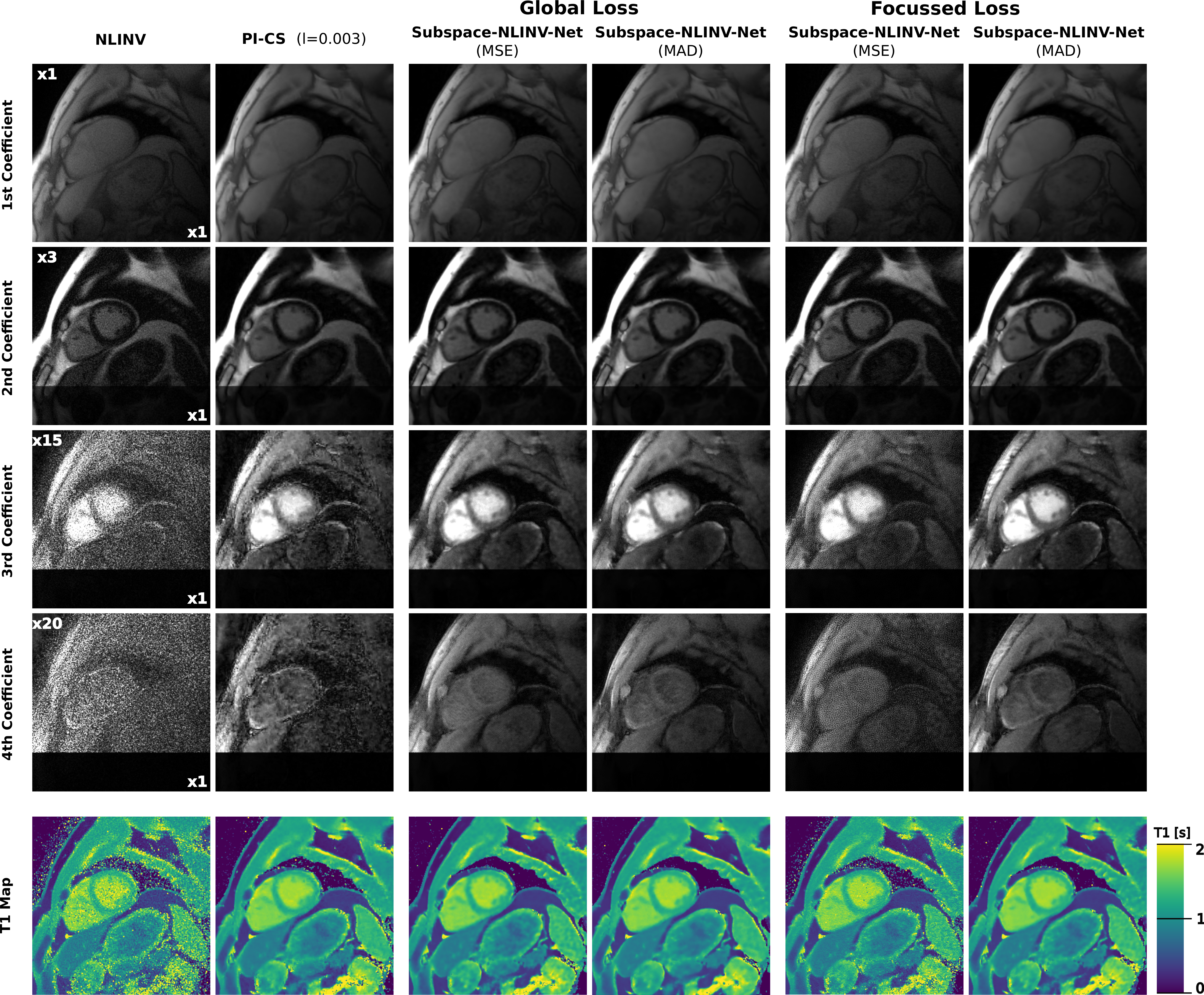}
	\caption{Subspace reconstruction and corresponding $T_1$-maps of a (mid) short-axis slice with ROVir-CC using NLINV, an $\ell^1$-regularized PI-CS reconstruction and Subspace-NLINV-Net reconstructions trained by different losses.
	Compared to SVD-CC in \figref{fig:coeff_svd}, streakings from outside the FOV are suppressed.
	}
	\label{fig:coeff_rovir}
\end{figure*}

We show coefficient maps and post-processed T1-maps using NLINV, PI-CS, and NLINV-Net in Figures \ref{fig:coeff_svd} (SVD-CC) and \ref{fig:coeff_rovir} (ROVir-CC).
For SVD-CC in \figref{fig:coeff_svd}, PI-CS and NLINV-Net denoise the coefficient maps, especially in the important region of the heart, while noise remains in the bottom-right corner of the third and fourth coefficient of the PI-CS reconstruction.
This leads to denoised post-processed T1-maps.
Streaking artifacts originating from outside the FOV are visible in the bottom-right corner of the NLINV-Net reconstructions, while they are covered by noise in the NLINV and PI-CS reconstructions.
These streakings are also visible in the corresponding T1 maps.
The reconstruction of NLINV-Net trained with the focussed MSE loss is corrupted by a high-frequency pattern, which is especially visible in the third and fourth coefficient map.
Considering the final T1-maps, NLINV-Net trained with MAD loss or global MSE loss functions performs similarly while providing slightly sharper edges at the papillary muscles and myocardial wall compared to the PI-CS reconstruction.
The corresponding reconstructions using ROVir-CC are presented in \figref{fig:coeff_rovir}.
The main difference to SVD-CC is that the streakings from outside the FOV are suppressed by ROVir-CC in all reconstructions.
Additionally, the remaining noise in the third and fourth coefficient map of the SVD-CC PI-CS reconstruction is suppressed in the ROVir-CC PI-CS reconstruction.

\begin{figure*}
	\includegraphics[width=\linewidth]{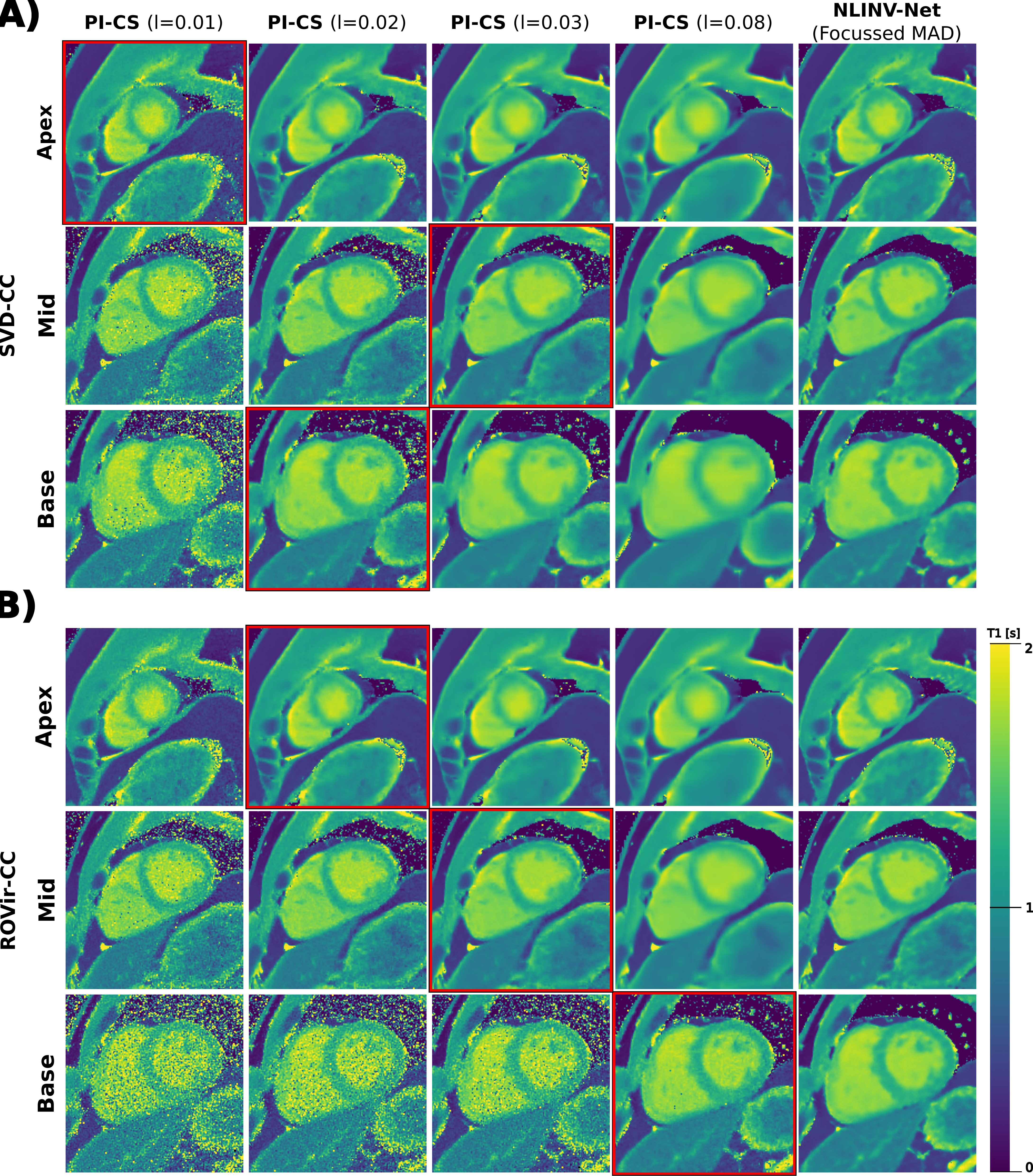}
	\caption{
		Reconstructed $T_1$-maps of one volunteer for PI-CS reconstruction with different regularization strength and NLINV-Net using SVD-CC (A) and ROVir-CC (B).
		Visually optimal regularization strength is highlighted by red boxes.
	}
	\label{fig:T1_maps}
\end{figure*}

\figref{fig:T1_maps} shows the three T1-maps of one volunteer for SVD-CC (A) and ROVir-CC (B) reconstructed by PI-CS with various regularization strength and NLINV-Net trained by the focussed MAD loss.
All images are cropped to 160x160 pixels which contain the heart in its center. 
It can be observed that the optimal regularization strength for PI-CS depends on the slice and the coil compression method, while the NLINV-Net reconstructions work well on all slices.
Overall, the NLINV-Net reconstructions remove slightly more noise from the reconstruction compared to the optimal PI-CS reconstructions, while not oversmoothing it.

\section{Discussion}

In this paper, we describe the use of NLINV-Net with SSDU training as a fully self-contained method that jointly
estimates image content and coil-sensitivity maps from radial MRI data.

In the two considered challenging applications, real-time cardiac imaging and cardiac single-shot T1 mapping, where ground
truth data is not available, NLINV-Net significantly denoises the reconstruction compared to plain (RT-)NLINV.
For the real-time dataset, most typically used regularization methods for PI-CS reconstructions such as temporal
total variation are not applicable if the reconstruction should be performed online. For this use case, RT-NLINV 
provides a causal and RT-NLINV-Net an almost causal reconstruction with a delay of only three frames.
The current implementation of RT-NLINV-Net is optimized for memory-efficient training and not for online reconstruction such that the current reconstruction is about 12-fold slower than the acquisition.
Ignoring CPU overhead, the computation time on the GPU measured with Nvidia Nsight Systems is two times slower than the acquisition.
By further optimization we expect to reduce the reconstruction time below the acquisition time similar to our previous work \cite{Uecker_NMRBiomed._2010}.

The reconstruction of the T1 datasets allows for a comparison with a state-of-the-art PI-CS reconstruction using joint $\ell^1$-Wavelet regularization.
While Subspace-NLINV-Net provides slightly sharper reconstructions as the $\ell^1$-Wavelet PI-CS method, the main advantage is that after training 
all parameters of the Subspace-NLINV-Net reconstruction are optimally tuned.
In contrast, the quality of the reconstruction with PI-CS depends on the chosen regularization parameter, where
the optimal regularization parameter may depend on many factors, such as the slice and coil compression method.

In all our experiments, NLINV-Net trained with the MAD loss performed superior to the counterpart trained with the MSE loss.
Considering the signal model (\eqref{eq:signal}), this is counterintuitive. From a statistical point of view, the maximum likelihood estimator for the Gaussian noise corresponds to optimizing the MSE loss.
On the other hand, the MAD loss is more robust against outliers in the k-space data, which may also originate from model imperfections in the signal model.
Typical imperfections in the setting of radial cardiac MRI are gradient delays or other gradient imperfections or off-resonance effects such as chemical shift
and field inhomogeneities.
If a neural network is trained in a supervised way, the objective is to recover the reconstruction corresponding to a reference reconstruction.
Hence, the network can learn how to deviate from the signal model to better match this reference reconstruction.
This is not possible in the SSDU framework as the signal model is the basis of the training objective, which underlines the importance of accurate modeling of the signal model.
In this work, we improved correspondence to the model in the loss by suppressing signal from peripheral regions using the ROVir coil compression method. 
While the selection of the region of interest was quite rough based on a fixed region of the FOV for all datasets, the method improved classical and NLINV-Net reconstructions.
Defining more fine-tuned regions of interests and sources of model errors such as vessels - potentially based on automatic segmentations - might be an interesting topic for further investigation.

While suppressing unwanted signal was a key objective of the development of ROVir-CC we extended its use to define a focussed loss function for SSDU.
The first virtual coil of ROVir-CC effectively suppresses signal from selected regions.
By restricting the self-supervised training loss to this virtual coil, we can suppress disturbing signal components from the periphery.
Interestingly, we observed that the focussed loss does not improve the reconstruction of the T1 dataset and if Subspace-NLINV-Net is trained with a focussed MSE loss, the reconstruction is even worse than with a global MSE loss.
In contrast, we observed improvements due to the focussed loss in the reconstructions of the real-time reconstruction.
Potentially, this is because the network can learn the cardiac motion better, which is mostly restricted to the focussed region.

Finally, we want to highlight the flexibility of the proposed network structure of NLINV-Net.
The joint estimation of coil sensitivity maps and image reconstruction make it  directly applicable where other reconstruction methods
may not work well because of errors in the estimation of coil sensitivity maps 
from non-Cartesian auto calibration regions.
Here, the integrated end-to-end implementation of NLINV-Net simplifies its application.

\section{Conclusion}

In this work, we proposed NLINV-Net, a new neural network architecture for
image reconstruction that can be trained directly on non-Cartesian data
using the SSDU training strategy without the need for ground truth 
reconstruction or pre-estimated coil sensitivities. We showed feasibility
and improved results for two challenging applications: real-time cardiac imaging
and single-shot T1 mapping.

\section*{Conflict of Interest}
The authors declare no competing interests.

\section*{Data Availability Statement}
In the spirit of reproducible research, the code to reproduce the results of this paper is available at \url{https://github.com/mrirecon/nlinv-net} (Version v0.3). All reconstructions have been performed with BART \cite{bart__2023}. Pretrained network weights are available at Zenodo \doi{10.5281/zenodo.11469859}.
BART is available at \url{https://github.com/mrirecon/bart}.
The data used in this study is available at Zenodo \doi{10.5281/zenodo.10492333} for RT-NLINV-Net and \doi{10.5281/zenodo.10491940} for Subspace-NLINV-Net.

\section*{Acknowledgements}
\printfunding
The authors thank the ISMRM Reproducible Research Study Group for conducting a code review of the code (Version v0.1) supplied in the Data Availability Statement. The scope of the code review covered only the code's ease of download, quality of documentation, and ability to run, but did not consider scientific accuracy or code efficiency.

\printbibliography

\clearpage
\appendix
\section*{Supporting Information}

Supporting Videos are available at Zenodo \doi{10.5281/zenodo.10845041}.

\subsection*{Video S1}
Video showing the real-time reconstruction of RT-NLINV and RT-NLINV-Net trained with a global and focussed MAD loss. The data corresponds to the data shown in \figref{fig:rt1}.

\subsection*{Video S2}
Video showing the real-time reconstruction of RT-NLINV and RT-NLINV-Net trained with a global and focussed MAD loss. The data corresponds to the data shown in \figref{fig:rt2}.

\subsection*{Video S3}
Video showing the post processed real-time reconstruction as in \figref[A]{fig:rt_post}.

\subsection*{Video S4}
Video showing the post processed real-time reconstruction as in \figref[B]{fig:rt_post}.

\subsection*{Video S5}
Video showing the coil sensitivities of the time series of the reconstruction in \figref{fig:rt1} with ROVir-CC. The color encoding visualizes the phase of the coil sensitivities and the reconstruction.

\subsection*{Video S6}
Video showing the coil sensitivities of the time series of the reconstruction in \figref{fig:rt2} with ROVir-CC. The color encoding visualizes the phase of the coil sensitivities and the reconstruction.

\section*{Supporting Figures}

\begin{figure}
\includegraphics[width=\linewidth]{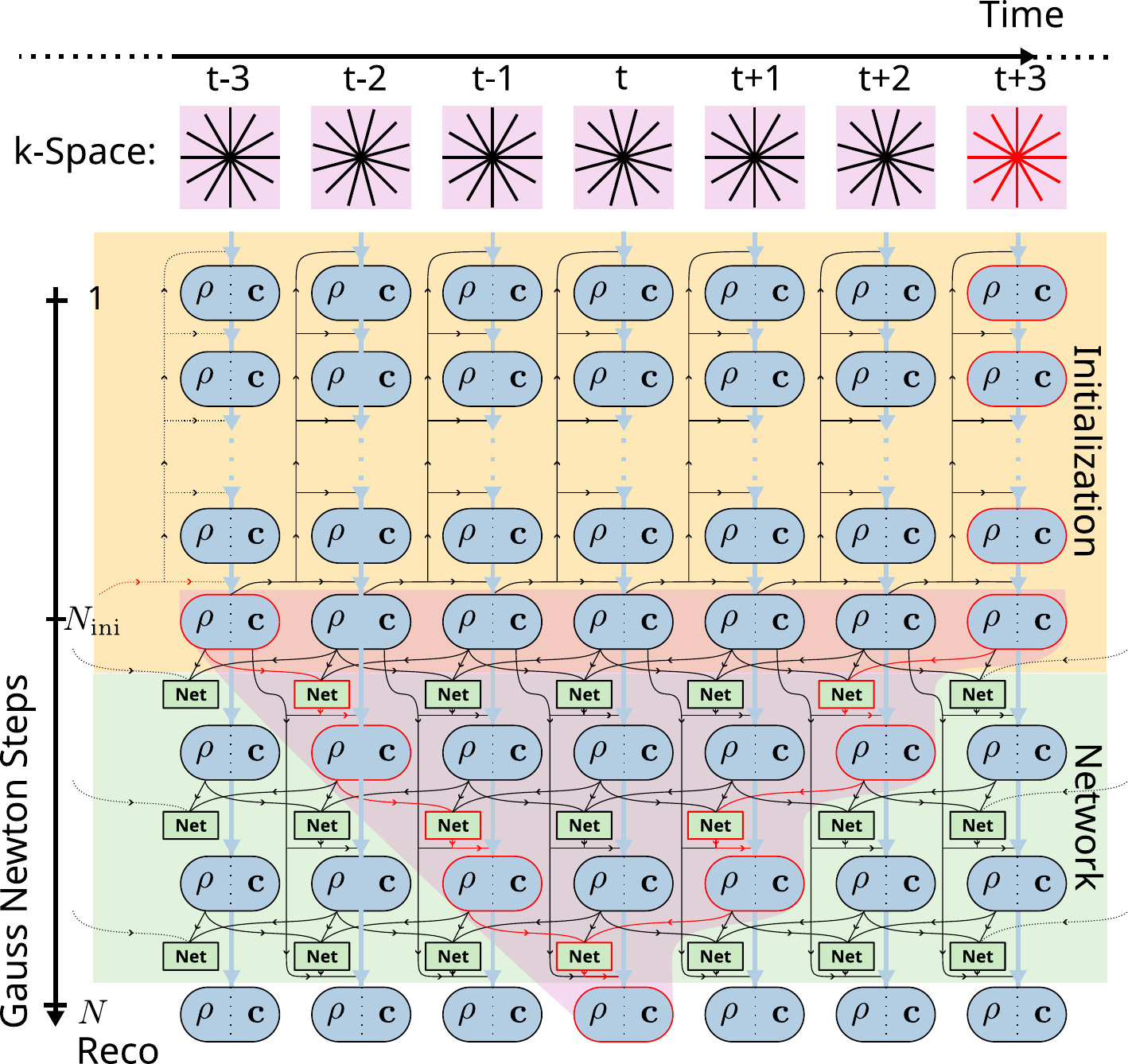}
\caption{Information flow in RT-NLINV-Net: Blue arrows visualize Gauss-Newton steps, while thinner black or red arrows visualize the dependence on reconstructions of previous frames or Gauss-Newton steps. The red arrows indicate the border of the receptive field of the reconstruction at time $t$.}
\label{supfig:receptive_field}
\end{figure}

\begin{figure}
\includegraphics[width=\linewidth]{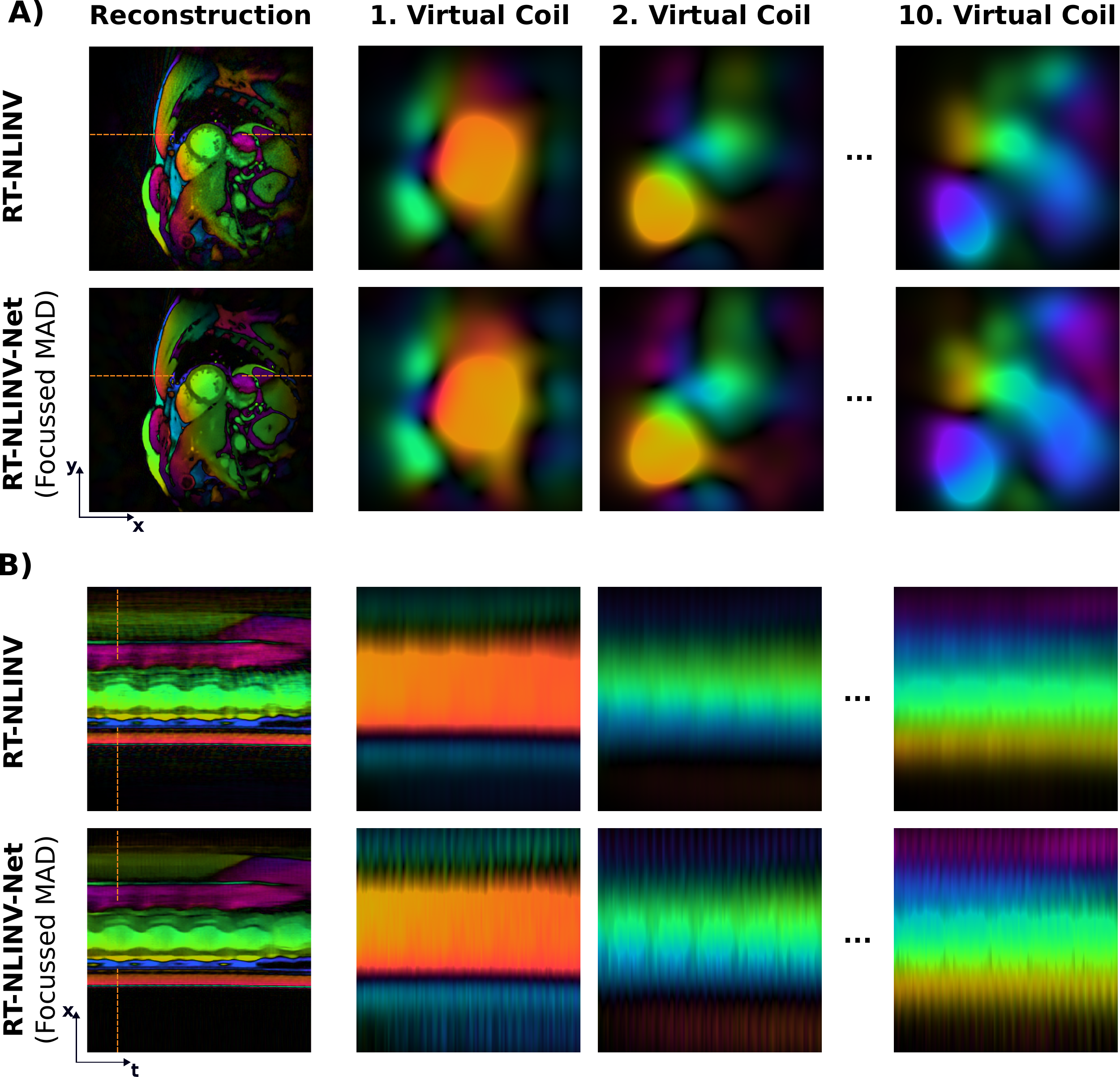}
\caption{Coil sensitivities and their time series for the reconstruction in \figref{fig:rt1} with ROVir-CC on the reconstruction FOV. The color encoding visualizes the phase of the coil sensitivities and the reconstruction.}
\end{figure}

\begin{figure}
\includegraphics[width=\linewidth]{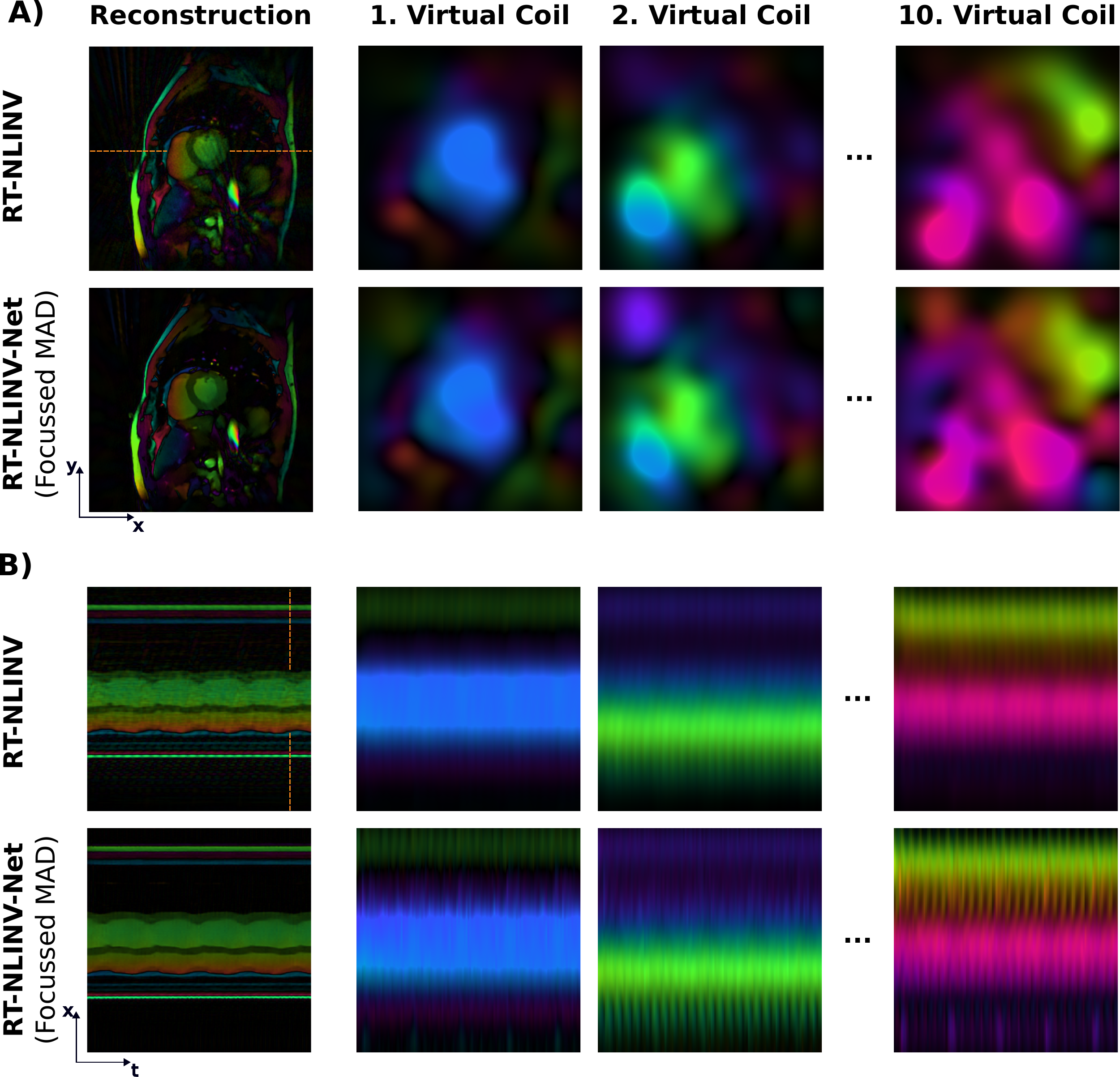}
\caption{Coil sensitivities  and their time series for the reconstruction in \figref{fig:rt2} with ROVir-CC on the reconstruction FOV. The color encoding visualizes the phase of the coil sensitivities and the reconstruction.}
\end{figure}

\begin{figure}
	\includegraphics[width=\linewidth]{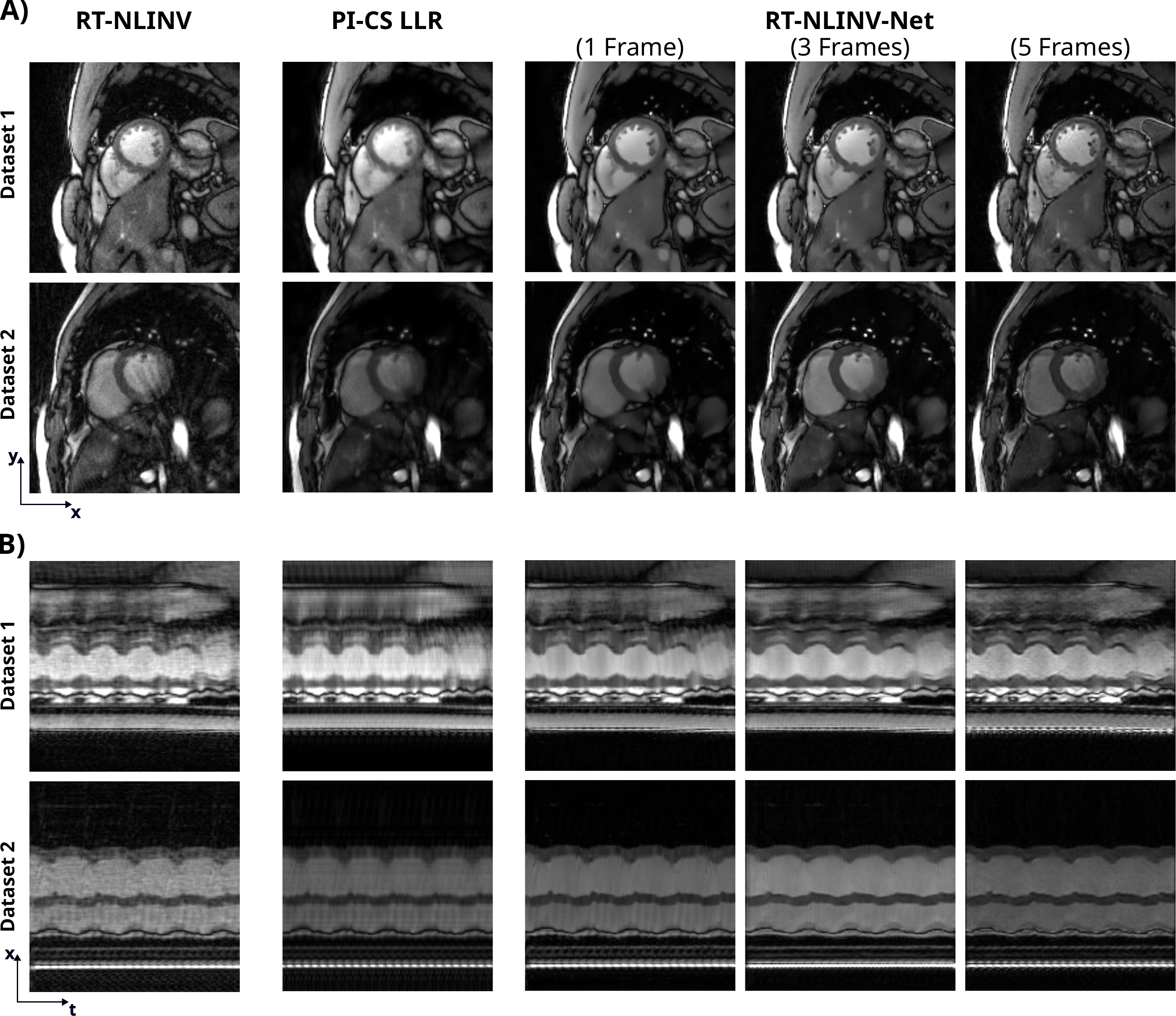}
	\caption{Comparison of RT-NLINV-Net with $N_F=1,3,5$ frames as input for the ResNet block and a locally low-rank (LLR) PI-CS reconstruction for the same datasets as shown in Figs. 3 and 4. The LLR-PI-CS reconstruction does not perform better than RT-NLINV. RT-NLINV-Net with a receptive field of $N_F=5$ frames slightly improves details compared to $N_F=3$ frames but this causes longer delay of the reconstruction in real-time imaging.}
	\label{supfig:llr_frames}
\end{figure}

\end{document}